\documentclass[fl eqn,twoside]{article}
\usepackage{amsmath}
\usepackage{amsfonts}
\usepackage{amssymb}
\def\be{\begin{equation}}

\def\ee{\end{equation}}

\def\bi{\bibitem}

\begin{document}

\title{Partition function of massless scalar field in Schwarzschild background.}

\author{Abhik Kumar Sanyal\\
Dept. of Physics, Jangipur College, Murshidabad, India - 742213}

\maketitle
\begin{abstract}
\noindent Using thermal value of zeta function instead of zero temperature, the partition function of quantized fields in arbitrary stationary backgrounds was found to be independent of undetermined regularization constant in even-dimension and the long drawn problem associated with the trace anomaly effect had been removed. Here, we explicitly calculate the expression for the coincidence limit so that the technique may be applied in some specific problems. A particular problem dealt with here is to calculate the partition function of massless scalar field in Schwarzschild background.
\end{abstract}

\footnotetext[1]{
\noindent Electronic address:\\
sanyal\_ ak@yahoo.com\\}

\section{Introduction}

We are now familiar with the fact that gauge-invariant divergences make general theory of relativity non-renormalizable. Nevertheless, the action containing higher order curvature invariant terms in the form,
\[A = \int\sqrt{-g} d^4 x \left[\frac{R - 2\Lambda}{16\pi G} + \alpha R^2 + \beta R_{\mu\nu}R^{\mu\nu}\right]\]
is renormalizable \cite{U-D}, even while interacting with matter \cite{Stelle} and also asymptotically free in all essential coupling constants \cite{Tomboulis1}-\cite{Tomboulis5}, so that there exists a truly cutoff-independent, interacting continuum field theory in $4$-dimension. However, analysis of linearized radiation shows that there are eight dynamical degree of freedom in the field. Apart from two massless spin-2 gravitons and a massive scalar particle, there are five massive spin-2 particles. Linearized field energies of the massive spin-2 particles excitations are negative definite. It is possible to recast the theory such that massive spin-2 eigenstates of the free field Hamiltonian have positive definite energy. But such an attempt results in negative norm in the state vector space, meaning that these are ghosts degrees of freedom and hence destroy the unitarity of the S-matrix. It is hoped that a modified version of string theory might turn out to be a complete theory of all interactions including gravity at and below the Planck scale. \\

However, presently, in the absence of a consistent quantum theory of gravity, one can study, as the first approximation, the quantum dynamics of the matter field in the background of classical gravity, which is dubbed as ``Quantum Field Theory in Curved Space Time". Quantum field theory in curved space-time is invariably plagued with divergences even in one loop order \cite{BD}. In fact, quantum field theory has divergences even in flat space-time and in all loop orders, including the first one. This results in the divergence of the partition function $Z \sim e^{S_S}$, $S_S$ being the classical action evaluated in Schwarzchild space. To get round this difficulty, one usually employs some sort of regularization technique. Unfortunately, such regularization techniques introduce new unknowns into the theory in the form of the so-called `regularization constants'. Unless one is fortunate enough to have a field for which this undetermined regularization constant can be taken care of by renormalization, the predictive power of the theory fall short of appealing to the experiments and there will be no way to fix the regularization constants. If the divergences occurring in two-loops and higher orders lead to the appearance of a finite number of `regularization constants' that could be taken care of by renormalization, the theory is said to be renormalizable. \\

It is therefore apparent that, one of the important problems of quantum field theory is the problem of renormalization. For the purpose, it is required to introduce external-field dependent counter-terms with dimension four. It was noted that these counter-terms are formed into geometrical invariants $R^2$ and $R^2_{\mu\nu}$ along with Einstein-Hilbert and cosmological terms \cite{U-D, U}. Using one-loop calculations in weak gravitational-field \cite{FF1}-\cite{FF3}, it was shown that to renormalize the theory of scalar field in curved space-time, renormalization constants of flat space are not sufficient. One new renormalization constant is needed, and it is necessary to introduce the $R\phi^2$ counter-term corresponding to nonminimal interaction of the scalar field with the gravitational field. One-loop counter-terms and renormalization constants in scalar field theories in external gravitational field with constant curvature had also been investigated by some authors \cite{count1}-\cite{count6}. The results of all the calculations had been presented in a nice review by Buchbinder, Odintsov and Shapiro \cite{BOS1, BOS2}. In a nut-shell, if the theory is renormalized in fiat space, then it could be only multiplicatively renormalized in curved space-time. This means that in curved space-time it is required to add external field Lagrangian with bare parameters to the bare Lagrangian which are renormalized by means of new renormalization constants. It is also required to add the terms corresponding to nonminimal interaction of matter with the gravitational field and renormalization of parameters of this nonminimal interaction is performed by means of new renormalization constants. The action for $R^2$ gravity presented above is one such multiplicative renormalization quantum theory that admits the renormalization group, although it is not unitary within the perturbation theory.
It is therefore clear that even the nonminimally coupled gravitational action

\[S = \int d^4 x\sqrt{-g}\left[\frac{1}{2}\xi R\phi^2 - \frac{1}{2}\phi_{,\mu}\phi^{,\mu} - \frac{1}{2}m^2\phi^2 - \frac{f}{4!}\phi^4\right]\]
in the absence of higher order curvature invariant term is not multiplicatively renormalizable \cite{BOS1}. Although the above action has the wonderful feature that it is generally covariant and also conformally invariant for $m = 0,~\xi = 1/6$. In view of above discussion, it is clear that a general quantum theory of gravity is non-renormalizable in some sense or other, though it is hoped that if better technique for summing the whole quantum perturbation series could be found, quantum gravity would turn out to be finite.\\

In the above context, there exists a reassuring result that, at least to one-loop order finite (non-zero) temperature field theory is free from the difficulties mentioned above. This had been pointed out long ago by Bandyopadhyay \cite{1N}, which went unnoticed. There, it was shown that after quantization, the partition function of a quantized field in arbitrary stationary curved space-time is free from the presence of any regularization constant which is in clear contradiction with the then known results \cite{book}. Indeed, earlier Hawking \cite{2H} had noted that the regularization constant does not appear in the partition function of a thermal massless scalar field in flat space-time. Nevertheless, it had been shown by Bandyopadhyay \cite{1N} that this fact is not particular to the massless scalar field in flat space-time but applies to quite general fields in arbitrary stationary backgrounds. A related fact of some importance was shown in the context of the absence of the so-called trace anomaly effect in the trace of the energy-momentum tensor of thermal massless fields in arbitrary stationary backgrounds. For this purpose a particular regularization technique, viz., the thermal value of zeta function was used (finite temperature theory) instead of zero temperature zeta function on a scalar field, in such a way that the partition function becomes finite in even space-time dimension  \cite{1N}. The method consists of compactification of time and Poisson re-summation. The result obtained by Bandyopadhyay \cite{1N} clearly contradicts the known result which states that ``the ultraviolet divergences of free energy do not depend on temperature'' \cite{Dowker}. Later, similar result was also found by Gusev and Zelnikov, although the authors did not interpret it as ``no-anomaly" statement \cite{gusev}.\\

The technique of zeta-function regularization was initiated by Dowker and Critchley \cite{DC}. Later, it was used for the calculation of one-loop effective potential in $f\phi^4$-theory and scalar electrodynamics on the de Sitter background \cite{zeta}. This technique had also been used for calculation of effective potential in more complicated theories on the de Sitter space \cite{zeta2}-\cite{zeta6}. A general technique of calculating one-loop effective action in De Sitter space by means of generalized zeta-function had also been presented \cite{zeta7}-\cite{zeta8}. Using Dimensional regularization implemented with zeta-function techniques some authors \cite{BEZ} had calculated the first quantum correction to the partition function at finite temperature for massless scalar fields on flat manifolds with non compact and with compact non-commutative dimensions. The partition function for the Schwarzschild - (anti-)de Sitter space has been evaluated by the WKB approximation also for $F(R)$ theory of gravity \cite{Cog}. However, $R_{\mu\nu}R^{\mu\nu}$ term has not been taken into account and therefore the Lagrangian is non-renomalizable \cite{Cog}. Here, the calculation of the partition function in Schwarzschild background for a renormalized theory of gravity has been posed. In section 2 we briefly review the results obtained by Bandyopadhyay \cite{1N} in a slightly different manner. In section 3, we have explicitly evaluated the coincidence limit with an eye to the possible applications of the results in specific problems. In section 4 we show the importance of deriving the results in section 3 by calculating the partition function of the massless scalar field in Schwarzschild background.\\

\section{Partition function of a quantized field in an arbitrary stationary background.}

For a multi-component quantized field $\phi^k$ in thermal equilibrium at temperature $T$ in a background of $n-$dimensional space with Lorentzian metric signature $(-,+,+,+,...,+)$, the partition function $Z$ is given by \cite{2H, 3H}

\be Z = N\int e^{-S_E[\Phi]}d\Phi,\ee
where, $S_E[\Phi]$ is the Euclidean action functional of the field obtained from the Lorentzian action functional $S_L[\Phi]$ under Wick rotation viz, replacing $t$ by $-i\tau$ and setting $S_E[\Phi] = -i S_L[\Phi]$. The transformed $n-$dimensional space has a positive definite metric with signature $(+,+,+,+,....,+)$. In expression (1) $d\Phi$ stands for $\Pi_k[d\phi^k]$, where, $k$ plays the role of both the discrete label characterizing the field components and the continuous label characterizing the space points. The $\tau (= x^0)$ integration in $S_E[\Phi]$ is over the interval $0\le\tau\le\beta$, where $\beta = \frac{1}{T}$ under the choice that the Boltzmann constant $\mathrm K_B = 1$ and the $\Phi$ integration in expression (1) is only over fields that are periodic in $\tau$ with period $\beta$. $N$ is the normalization constant.\\

\noindent
For the sake of simplicity it will be assumed that all the field components are bosonic and that invariant flows, arising out of the gauge invariance of the action are absent. It is important to note that these assumptions are not essential for the conclusions drawn in \cite{1N}. When the action functional is not quadratic in the fields, the standard procedure to obtain the one-loop contribution to the partition function $Z$ is to expand $S_{E}[\Phi]$ about a background field $\Phi_0$ and retain upto terms quadratic in the field. In the process, one-loop order $Z$ is given by

\be Z=N e^{-S_E[\Phi_0]}\int e^{-\frac{1}{2}\tilde\Phi {\mathcal{F}}\Phi}d\Phi,\ee
where, $\tilde\Phi$ stands for the transpose of the multi-component quantity ($\Phi - \Phi_0$) and

\be {\mathcal{F}}_{ij} = \frac{\delta^2 S_E[\Phi]}{\delta\Phi^i\delta\Phi^j}\Big|_{\Phi = \Phi_0}.\ee
The Gaussian integral in expression (2) can be expressed in terms of the determinant of the operator $\widehat{F} = g^{-\frac{1}{4}} {\mathcal F} g^{-\frac{1}{4}}$, where, $g = \mathrm{det} (g_{\mu\nu})$ and $g^{-\frac{1}{4}}$ has been inserted to render the integrand (2) generally covariant. Thus, the partition function is expressed as (see Appendix 7.1)

\be \ln Z = - S_{E}[\phi_0] + \int \omega(\vec x)d^n x + \mathrm {constant},\ee
where the one-loop Euclidean effective action is given by
\be \omega(\vec x) = \frac{1}{2}\int \mathrm{Tr} [K(\vec x, \vec x', s)]\frac{d s}{s}.\ee
It is not difficult to see that the expression (4) is actually divergent. To make $\ln Z$ finite, Bandyopadhyay \cite{1N} followed $\zeta$-function regularization procedure. The generalized $\zeta$-function corresponding to $\widehat{F}$ is defined as,

\be \zeta(\vec x, Z) = \sum_p\sum_\alpha \frac{v(\vec x, p, \alpha) v^\dag(\vec x, p, \alpha)}{\left[\mu^{-2} \lambda(p, \alpha)\right]^Z}\ee
while
\be \zeta(Z) = \int \mathrm{Tr} \zeta(\vec x, Z)d^n x = \sum_p\sum_\alpha \frac{\mu^{2Z}}{[\lambda (p, \alpha)]^Z}.\ee
More precisely, $\zeta(\vec x, Z)$ is defined by the above relation, first in the region of the complex $Z$-plane where the sum converges and then it is defined elsewhere by analytic continuation. The regularization constant $\mu$ is introduced so that $\zeta(Z)$ is dimensionless and the dimensions of $\zeta(\vec x, Z)$ remain fixed under the analytic continuation. Now, one can observe that

\be \zeta(\vec x, Z) = \frac{1}{\Gamma(Z)}\int_0^\infty \left(\mu^2 s\right)^{Z-1} K(\vec x, \vec x', s) \mu^2 ds\ee
In view of expression (7) one can formally write

\be \det\left(\mu^2 s\right) = e^{-\zeta'(0)},\ee
and obtain the expression for $\omega(\vec x)$ as

\be \omega(\vec x) = \frac{1}{2} \mathrm{Tr} \frac{\partial}{\partial Z} \zeta(\vec x, Z)\Big|_{Z = 0}.\ee
In the process the `regularized' value of $\ln Z$ presented in equation (4) is now obtained. At this end, for the sake of illustration let us now consider the $r-$component scalar field $\Phi^i ~(i = 1,2,...,r)$ in an arbitrary stationary curved space-time, for which the Euclidean action reads

\be S_{E}[\Phi] = \int \left[\sqrt g~ g^{\mu\nu} \tilde{\Phi}_{,\mu}\tilde{\Phi}_{,\nu} + (m^2 + \xi R)\tilde{\Phi}\Phi\right]d^n x,\ee
suppresseing the components. In the above, $\tilde{\Phi}$ stands for the transpose of the multi-component scalar field. It is apparent that the kernel $K(\vec x, \vec x', s)$ is periodic in ($\tau - \tau'$) with period $\beta$ (see Appendix) and therefore one can set

\be K(\vec x, \vec x', s) = \frac{1}{\beta}\sum_{N = -\infty}^{\infty} \frac{e^{i\frac{2\pi}{\beta}N(\tau - \tau')}}{(4\pi s)^{\frac{n-1}{2}}}\sqrt D~e^{-\frac{\sigma(\vec x, \vec x')}{2s}-\left(m^2 + \frac{4\pi^2 N^2}{{\mathcal N}^2\beta^2}\right)s}\Lambda_{N}(\vec x, \vec x', s).\ee
In the above expression $\sigma(\vec x, \vec x')$ is the half square of the geodesic distance between $\vec x$ and $\vec x'$ in the $(n-1)$ dimensional space, ${\mathcal N}$ is the lapse function of the $n-$dimensional Euclidean space-time, $\Lambda_{N}(\vec x, \vec x', s)$ are some $r \times r$ matrices to be determined under the condition

\be \Lambda_{N}(\vec x, \vec x', 0) = 1_r ~\mathrm{for ~each}~ N,\ee
and $D = \det (-\sigma_{;ij})$, where semicolon denotes covariant derivative with respect to the space part of the metric $(g_{ij} = \gamma_{ij})$. It is therefore obvious that

\be \lim_{\vec x' \rightarrow \vec x} D(\vec x, \vec x') = \gamma = \det(\gamma_{ij}).\ee
Now substituting the expression of the kernel (12) in expression (8), one obtains

\be \zeta(\vec x, Z) = \frac{1}{\beta}\sum_{N = -\infty}^{\infty} \frac{\mu^{n-1}\sqrt\gamma}{(4\pi )^{\frac{n-1}{2}}\Gamma(Z)}\int_0^{\infty}(\mu^2 s)^{Z-1-{n-1\over 2}} e^{-\left(m^2 + \frac{4\pi^2 N^2}{{\mathcal N}^2\beta^2}\right)s}\Lambda_{N}\mu^2 d s,\ee
since at $x\rightarrow x'$, $\tau\rightarrow \tau'$ and so, $e^{{i 2\pi\over \beta}N(\tau-\tau')} = 1$. Now remembering the fact that $\lim_{s\rightarrow\infty} e^{-im^2 s}\Lambda(\vec x,\vec x',s) = 0$, one can integrate the above equation by parts for ${Z} > {n-1\over 2}$ to obtain

\be \zeta(\vec x, Z) = \frac{1}{\beta}\sum_{N = -\infty}^{\infty} \frac{\mu^{n-1}\sqrt\gamma}{(4\pi )^{\frac{n-1}{2}}(Z - \frac{n-1}{2})\Gamma(Z)} \left[
-\int_0^{\infty}\frac{\partial}{\partial s} \left\{e^{-\left(m^2 + \frac{4\pi^2 N^2}{{\mathcal N}^2\beta^2}\right)s}\Lambda_{N}\right\}
(\mu^2 s)^{Z-\frac{n-1}{2}}ds\right]\ee
since the first term vanishes at the end points. Integrating the above equation by parts yet again and getting rid of the first term which vanishes at the end points, one finally obtains

\be \zeta(\vec x, Z) = \frac{1}{\beta}\sum_{N = -\infty}^{\infty} \frac{\mu^{n-3}\sqrt\gamma}{(4\pi )^{\frac{n-1}{2}}(Z - \frac{n-1}{2})(Z - \frac{n-3}{2})\Gamma(Z)} \left[\int_0^{\infty}
(\mu^2 s)^{Z - \frac{n-3}{2}}\frac{\partial^2}{\partial s^2} \left\{e^{-\left(m^2 + \frac{4\pi^2 N^2}{{\mathcal N}^2\beta^2}\right)s}\Lambda_{N}\right\}
ds\right].\ee
Now inspecting the above equation in different dimensions, even or odd, corresponding to different values of $n$, one finds the following general expression (see Appendix 7.2)

\be \zeta(\vec x, 0) = \left\{ \begin{array}{ll}
                      0  & ,\mathrm {for}~ n~\mathrm {even} \\

                      \frac{1}{\beta} \sum_N \frac{\sqrt{\gamma}~(\vec x)}{\left(\frac{n-1}{2}\right)!(4\pi)^{\frac{n-1}{2}}}\Big[\left(\frac{\partial}{\partial s}\right)^{\frac{n-1}{2}}e^{-\left(m^2 + \frac{4\pi^2 N^2}{{\mathcal N}^2 \beta^2}\right)s}\Lambda_N(\vec x, \vec x', s) \Big] & ,\mathrm {for}~n~\mathrm {odd} \end{array} \right. \ee
The one-loop Euclidean effective action may now be calculated in view of equation (10) to obtain (see Appendix 7.2)
\be \omega(\vec x) ~~~= \frac{1}{\beta} \sum_N \frac{\sqrt{\gamma}~(\vec x)}{12 \pi^{\frac{3}{2}}}
\int_0^\infty s^{-\frac{1}{2}}\left(\frac{\partial}{\partial s}\right)^2  e^{-\left(m^2 + \frac{4\pi^2 N^2}{{\mathcal N}^2 \beta^2}\right)s}\mathrm{Tr}[\Lambda_N(\vec x, \vec x', s)]d s ~~,~ (n  = 4).\ee
The important difference in results obtained by Bandyopadhyay \cite{1N} in comparison with that obtained by De-Witt \cite{4D} is now apparent. While the thermal zeta function at $Z = 0$ vanishes for even-dimensional space-time $n$ in Bandyopadhyay's calculation \cite{1N}, zero temperature zeta function at $Z = 0$ vanishes for odd $n$ as obtained by DeWitt \cite{4D}. Expression (9) in view of (19) clearly implies that in the physical four dimensional space-time $Z$ is independent of any undetermined regularization constant. Further, since the massless conformally invariant scalar field ($m = 0 ~\&~ \xi = \frac{(n-2)}{4(n-1)}= \frac{1}{6}$ in $4$ dimensional space-time) admits \cite{4D}

\be \big< T^\mu_{~\mu} \big> = -\mathrm{Tr}[\zeta(\vec x,0)],\ee
so expression (19) implies that in the physical four dimensional space $\big< T^\mu_{~\mu} \big>$ vanishes and therefore the so called trace anomaly effect is absent. Finally, the regularized value of $\ln Z$ is now obtained as

\be \ln Z = \frac{1}{\beta}\sum_{N=-\infty}^{+\infty} \frac{1}{12 \pi^{\frac{3}{2}}}
\int_0^\infty s^{-\frac{1}{2}}\left(\frac{\partial}{\partial s}\right)^2 \left[\int \sqrt{\gamma}~(\vec x) e^{-\left(m^2 + \frac{4\pi^2 N^2}{{\mathcal N}^2 \beta^2}\right)s}\mathrm{Tr}[\Lambda_N(\vec x, \vec x', s)]d^n x\right] d s,\ee
where, for spaces with non-trivial metric, the space integration has to be performed before the $s$-integration. The partition function (21) obtained by Bandyopadhyay \cite{1N} is of particular importance. Firstly, the partition function is found to be independent of the regularization constant $\mu$ in the $4-D$ space-time. This result is similar to the one, found later by Gusev and Zelnikov \cite{gusev}. Next, for massless conformally invariant scalar fields, in view of the expression (20), the trace of the energy-momentum tensor of thermal quantized massless scalar field $\big<T^{\mu}_{~\mu}\big>$ vanishes ie., trace anomaly is absent, in contrast to Hawking's result \cite{2H}. Finally, $\ln Z$ for massless scalar field in flat space-time calculated by Bandyopadhyay \cite{1N} agrees with the result obtained by Hawking \cite{2H}, who used zero temperature $\zeta$ function instead. Still the partition function obtained by Hawking was independent of the regularization constant $\mu$ and agrees with Bandyopadhyay \cite{1N} because, he calculated $\zeta$ function by summing over eigenvalues corresponding to periodic eigenfunctions only. Nevertheless, the general result that the partition function is independent of regularization constant and absence of trace anomaly suggests that regularization scheme should be followed with thermal $\zeta$ function. In the following sections we show that such a result enables one to calculate the partition function in more general situations, like massless scalar field in Schwarzschild background, which has not been attempted earlier due to the presence of undetermined regularization constant in non-flat background.

\section{Evaluation of the coincidence limits $\Lambda_N(\vec x, \vec x', 0).$}
In the previous section we have reviewed the technique of using thermal $\zeta$ function regularization scheme to obtain the expression (21) of partition function free from regularization constant, obtained by Bandyopadhyay \cite{1N}. In this context we hqave also discussed the main difference with the one obtained by Hawking using vacuum to vacuum transition amplitude instead. Now, to realize the utility of the expression  of $\ln Z$, one should try to apply the expression (21) in some specific problems. At the first place, it is therefore required to find explicit expression of the coincidence limit $\Lambda_N(\vec x, \vec x', 0)$, which we pose in this section. For this purpose, let us assume the power series expansion

\be \Lambda_N(\vec x, \vec x', s) = \sum_{q=0}^{\infty} a_{Nq}(\vec x, \vec x')s^q\ee
with
\be a_{N0}(\vec x, \vec x') = 1_r, ~\mathrm{for ~each}~N,\ee
where, $q$ is an integer. Also for static positive definite metrics of the form

\be ds^2 = {\mathcal{N}}^2 d\tau^2 + \gamma_{ij} dx^i dx^j\ee
one has
\be  \widehat{F}  = \gamma^{-1/2} \pi_i \gamma^{1/2} \gamma^{ij} \pi_j + f(\vec x) + m^2 - \frac{1}{{\mathcal{N}}^2}\partial_0^2\ee
where,
\be \pi_i = -i \left[\partial_i + (\ln {\mathcal{N}}^{1/2})_{;i}\right]\ee
and
\be f(\vec x) = (\ln {\mathcal{N}}^{1/2})^{;i}_{;i} + \gamma^{ij} \frac{{\mathcal{N}}_{,i}{\mathcal{N}}_{,j}}{4 {\mathcal{N}}^2} + \xi R. \ee
Now substituting the expression for the heat kernel (12) in the expression of the differential equation of the same (equation 49 of appendix 7.1) and using Expression (25) one obtains the following recursion relation for the $a_{Nq}$'s

\be \sigma^{;i}a_{N0;i} = 0,\ee
\be\begin{split} \sigma^{;i}a_{Nq;i} + p a_{Nq} &= D^{-1/2}\left(D^{1/2} a_{N, q-1}\right)^{;i}_{;i} + \frac{4\pi^2 N^2}{\beta^2} \sigma^{;i}\left(\frac{1}{{\mathcal N}^2}  \right)_{;i} a_{N, q-1}  \\
& - \frac{4\pi^2 N^2}{\beta^2} \left(\frac{1}{{\mathcal N}^2} \right)^{;i}_{;i} a_{N, q-2} - 2\frac{4\pi^2 N^2}{\beta^2} D^{-1/2}\left(D^{1/2} a_{N, q-2}\right)_{;i}\left(\frac{1}{{\mathcal N}^2}\right)^{;i} \\
& + \left(\frac{4\pi^2 N^2}{\beta^2}\right)^2\left(\frac{1}{{\mathcal N}^2} \right)_{;i} \left(\frac{1}{{\mathcal N}^2}\right)^{;i} a_{N, q-3} - f(\vec x)a_{N, q-1}.\end{split}\ee
Expressions (23), (28) and (29) can now be used to find the coincidence limits $a_{Nq}(\vec x, \vec x')$. For details of the procedure one may refer to  Dewitt \cite{4D}. Here we only state the results of our calculations, which are,

\be a_{N0}(\vec x, \vec x') = 1_r,\ee
\be a_{N1}(\vec x, \vec x') = \left[ \frac{1}{6} R - f(\vec x)\right] 1_r,\ee
\be a_{N2}(\vec x, \vec x') = \left[\frac{1}{6}\left(\frac{1}{5}R_{;i}^{;i}-f(\vec x)^{;i}_{;i}\right)+ \frac{1}{2}\left(\frac{1}{6}R -f(\vec x)\right)^2 - \frac{1}{180}\left(R_{ij}R^{ij}-R_{ijkl}R^{ijkl}\right) + \frac{4\pi^2 N^2}{\beta^2}\left(\frac{1}{{\mathcal N}^2}\right)^{;i}_{;i}\right]1_r,\ee
etc. In the above equations $R, R_{ij} {~\mathrm {and}~} R_{ijkl}$ stand for Ricci scalar, Ricci tensor and Riemann curvature tensor respectively of the $(n-1)$ dimensional space with metric $\gamma_{ij}$.

\section{Partition function ($\ln Z$) for one component massless scalar field in Schwarzschild background}

Having obtained explicit expression for the coincidence limit, we are now in a position to calculate the partition function corresponding to specific problems. Here, we find the partition function for massless scalar field in Schwarzschild background,

\be d S_E^2 = \left(1-\frac{2M}{r}\right)d\tau^2 +  \left(1-\frac{2M}{r}\right)^{-1}dr^2 + r^2 d\Omega^2,\ee

\noindent
retaining terms for the coincidence limit $a_{Nq}$ upto $a_{N1}$. On the basis of the so-called `scaling argument' Hawking \cite{2H} conjectured that the partition function for the quantized massless scalar field contained in a large spherical box of radius $r_0$ and in thermal equilibrium with Schwarzschild black hole of mass $M$ will be given by

\be \ln Z = \zeta(0)\ln\left(\frac{M}{M_0}\right) + f\left(\frac{r_0}{M}\right),\ee
where $\zeta(r)$ is the generalized zeta function of the scalar wave operator in the `Euclidean section' of the Schwarzschild space time, $M_0$ is a constant related to the `regularization constant' and $f(\frac{r_0}{M})$ is some function that is expected to have the form

\be f\left(\frac{r_0}{M}\right) = \frac{r_0^3}{34560 M^3} + {\mathcal{O}}\left(\frac{r_0^2}{M^2}\right),\ee
in order to agree with the expression for $\ln Z$ in flat spacetime in the limit of large $\frac{r_0}{M}$ \cite{1N,2H}. We have already mentioned that in the calculation of $\ln Z$ one requires `thermal' $\zeta$-function rather than `zero temperature' $\zeta$-function. Further, we have also shown that in even-dimensional space-time and hence in the physical four dimensional space-time the `thermal' $\zeta$-function vanishes at zero argument for quite general systems irrespective of the background metric. Therefore, in the expression (34) for $\ln Z$, the first term vanishes and $\ln Z$ for the massless scalar field in Schwarzschild background is expected to have the form

\be \ln Z = \frac{r_0^3}{34560 M^3} + {\mathcal{O}}\left(\frac{r_0^2}{M^2}\right).\ee
We now propose to find $\ln Z$ upto those lower order terms which diverge when $\frac{r_0}{M}$ tends to infinity. For the purpose first of all we notice that the Euclidean section of the background Schwarzschild space is periodic in the Euclidean $\tau$ co-ordinate with period $8\pi M$ and extends in the radial direction from $r = 2 M$ to $r = \infty$ \cite{3H}. Thus, $\beta = \frac{1}{T}$ in the expression (21) for $\ln Z$ is to be set equal to $8\pi M$ in order to establish thermal equilibrium of the scalar field quanta with the whole of mass $M$. For the scalar field contained in a spherical box of radius $r_0 > 2M$ and in thermal equilibrium with the hole, the contribution to the spacetime integral in expression (22) of the $a_{N0}$ term, under a change of variable $1-\frac{2M}{r} = (Y+1)^{-1}$, is given by

\be I_{N0} = \int_{\tau = 0}^{\beta}\int_{r = 2M}^{r_0}\int_{\theta = 0}^\pi\int_{\phi = 0}^{2\pi}\sqrt{\gamma} e^{-\frac{4\pi^2N^2}{{\mathcal N}^2\beta^2}s}d\tau dr d\theta d\phi = 32\pi M^3 \int_{Y_0}^\infty
\frac{(Y + 1)^{\frac{5}{2}}}{Y^4}e^{-K(Y+1)}d Y,\ee
where,
\be Y_0 = \frac{2M}{r_0}\left(1 - \frac{2M}{r_0}\right)^{-1}~~\mathrm{and}~~K = \frac{4\pi^2 N^2}{\beta^2}s.\ee
The integral in expression (37) vanishes at the upper limit. Since for $r_0 \gg 2M $, $Y_0 \ll 1$, so in evaluating the integral in the limit of large $r_0$, we may expand $(Y + 1)^{\frac{5}{2}}$ in the integrand in a power series in $Y$. In this way, $I_{N0}$ can be calculated to any desired accuracy (More explicit evaluation of the integral has been presented in Appendix 7.3). However, we exhibit only those terms in $I_{N0}$ that diverge in the limit $r_0 \rightarrow \infty$,

\be\begin{split} I_{N0} = 32\pi M^3 e^{-K(Y_0+1)} &\Bigg[\frac{1}{3 Y_0^3}+\left(\frac{5}{4}-\frac{K}{6}\right)\frac{1}{Y_0^2} +\left(\frac{15}{8}+\frac{K^2}{6}-\frac{5K}{4} \right)\frac{1}{Y_0} \\
&+ \left(\frac{K^3}{6}-\frac{5K^2}{4}+\frac{15}{8}K-\frac{5}{16} \right)\ln Y_0 + ...\Bigg].\end{split}\ee
The contribution to $\ln Z$ of all the $a_{N0}$ terms for all values of $N$ is given by

\be\begin{split} I_0 &= 2\sum_{N=1}^\infty\frac{1}{12 \pi^{3/2}}\left(\frac{2\pi|N|}{\beta}\right)^3\int_{0}^\infty K^{-1/2}\left(\frac{\partial}{\partial K}\right)^2 I_{N0}(K)dK\\
&= \frac{128 \pi^{\frac{5}{2}} M^3\zeta_R(-3)\Gamma_{\frac{1}{2}}}{9\beta^3 Y_0^3} \Bigg[(Y_0+1)^{\frac{3}{2}}+ \left(\frac{15}{4}(Y_0+1)^{\frac{3}{2}}+ \frac{3}{4}(Z_0+1)^{\frac{1}{2}}\right)Y_0\\
& + \left(\frac{45}{8}(Y_0+1)^{\frac{3}{2}}+\frac{45}{8}(Y_0+1)^{\frac{1}{2}}+\frac{3}{8}(Y_0+1)^{-\frac{1}{2}}\right)Y_0^2\\
& -\left(\frac{15}{16}(Y_0+1)^{\frac{3}{2}}+\frac{135}{16}(Y_0+1)^{\frac{1}{2}}+
\frac{45}{16}(Y_0+1)^{-\frac{1}{2}}-\frac{3}{16}(Y_0+1)^{-\frac{3}{2}} \right)Y_0^3 \ln Y_0 + .....\Bigg].\end{split}\ee
It can be easily verified that the contribution of $a_{N1}, a_{N2},....$ terms to $\ln Z$ do not diverge at $r_0 \rightarrow \infty$. Thus $\ln Z$ upto terms that diverge in the limit $r_0 \rightarrow \infty$ is given by

\be \ln Z = \frac{1}{34560}\left[\left(\frac{r_0}{M}\right)^3 + 6 \left(\frac{r_0}{M}\right)^2 + 36 \left(\frac{r_0}{M}\right) + 96 \ln \left(\frac{M}{r_0}\right) + .....\right],\ee
which is the final expression for the required partition function of a massless scalar field in Schwarzschild background. It is important to note that for the expression of partition function, we only required the conjecture (34) presented by Hawking \cite{2H}. Likewise, once a conjecture is given for massive scalar field, it is in principle possible to find the expression of the partition function, as well.

\section{The issue of stabilization}
Particle creation phenomena by external fields like constant electric, magnetic or by quantum fluctuation of the vacuum is a well-established fact. Likewise, creation and annihilation of virtual black holes are also possible from the quantum fluctuations of the gravitational field. This lead to some sort of instability of the field which is manifested only at one-loop level. In fact, information in this regard may be obtained by means of saddle point methods. If a lower state exists, it can be reached by quantum tunneling and a negative mode appears in Lichnerowicz operator.  Such instability leads to the spontaneous nucleation of black holes, which actually signals transition from false vacuum to the true. The energy at one-loop represents Casimir like energy and this essentially gives a measure of vacuum fluctuations.\\

To establish this fact, a subtraction procedure to avoid divergences \cite{jd1}-\cite{jd2} coming from boundaries is followed, which is related to the Casimir energy. The imposed by-hand cut-off of the UV divergence of the free gravitons had been successfully accomplished and Casimir energy were calculated in Schwarzschild \cite{g1}, Schwarzschild-anti de Sitter \cite{g2} and Schwarzschild-de Sitter backgrounds \cite{g3}. A clear signal of instability following the existence of one negative mode in the TT sector [transverse, traceless] when the saddle point approximation is considered, was found \cite{pg}-\cite{ms}. In particular, the instability has been related to the probability of creating a black hole pair \cite{rg1}-\cite{rg2}. However this particular result has been obtained by looking at the partition function and therefore with the introduction of an equilibrium temperature, that in the case of the pure Schwarzschild and flat metric can be imposed to be equal. On the other hand, a tunnelling process was also found earlier which involved a flat space with non-zero temperature \cite{dj}. Although in that case, the initial space might not represent a correct vacuum, the result reflected the stability of flat space under quantum fluctuation via nucleation of black holes. In fact in finite temperature theory it is possible to generate single black holes by quantum fluctuations. Likewise, here we have taken into account the finite temperature field theory and considered thermal zeta function instead of zero temperature one. Since the zeta function is regular at the origin, and the partition function is finite in even dimensional spacetime, the one loop divergences is removed naturally and the the logarithm of the partition function has be expressed in terms of the zeta function and its derivative evaluated at the origin. In this case the vacuum energy (Casimir energy) can be defined resorting to the usual thermodynamical relation,

\be < E > = -\lim_{\beta\rightarrow\infty}\frac{\partial(\log Z)}{\partial \beta},\ee
which in the present case is

\be <E> = \frac{1}{11520\beta}\left[\Big(\frac{r_0}{M}\Big)^3+4\Big(\frac{r_0}{ M}\Big)^2 +12\Big(\frac{r_0}{M}\Big) -32 +...\right]\ee

which automatically makes the Casimir energy finite without the requirement of UV cut-off by hand. The sign of vacuum energy is relevant for issues governing stabilization mechanism. Since vacuum energy here is independent of renormalization parameter $\mu$, so its sign is fixed once and forever. It is clear from the above expression that the vacuum energy is positive for $r_o > 2M$. Thus the result shows no signal of instability as in the flat case .

\section{Concluding remarks}

Using thermal $\zeta$ function regularization scheme, instead of zero temperature one (ie., the $\zeta$ value corresponding to vacuum to vacuum transition amplitude) Bandyopadhyay \cite{1N} had proved that the partition function for quite general type of fields in stationary backgrounds and in the presence of invariant flows is independent of undetermined regularization constant in even dimension and therefore in physical $4$-dimensional space-time. It is very important to remember that the formal identity $T^{\mu}_{~\mu} = 0$, valid for conformally invariant classical field theories was found to fail for quantized fields, which is the so called trace anomaly. However, thermal $\zeta$ function regularization scheme resulted in the absence of trace anomaly.\\

Therefore, at the first place, we have reviewed to revive this important piece of work, which went unnoticed for more than two decades. We have also made comparison with the one presented by Hawking \cite{2H}. Our main objective here is to expatiate the applicability of the result \cite{1N}. For this purpose, in section 3, we have calculated the expression for the coincidence limit following power series expansion. In section 4, we have found the partition function of a massless scalar field in Schwarzschild background, without suppressing any of the modes. It is well known that the main objective in statistical mechanics is to find the partition function ($Z$) of a system. Starting with the partition function, all thermodynamical quantities can be determined. Of particular importance are the Helmholtz free energy $F = -KT\ln Z$, the average energy $ \bar E = -{\partial\ln Z\over\partial\beta}$ and the entropy $S = K(\ln Z + \beta\bar E)$, which can now be trivially computed from expression (41). The technique may be followed also to find the partition function for a quantized massless scalar field contained in a large spherical box of radius $r_0$ and in thermal equilibrium with Kerr black hole of mass $M$ and angular momentum $L$, which would include shift vector as well, had there been a conjecture like expression (34) for the same. In fact, we would like to emphasize that, once a conjecture as in expression (34) is present, for say, massive scalar field, one can compute its partition function following the same technique, as well.

\section{Appendix}
\subsection{The expression (4) for $\ln Z$}
Expression (2) gives
\be \ln Z = S_{E}[\Phi_0] - \frac{1}{2}\ln\det (\mu^{-2} \widehat{F}) + \mathrm{constant}.\ee
In the above expression, the dimension of the constant $\mu$ is mass or length$^{-1}$, under the assumption that $\widehat{F}$ is a second order differential operator and it has been introduced to make the determinant dimensionless. It is important to note that since the field integration in expression (1) is over periodic fields only, $\det (\mu^{-2} \widehat{F})$ in expression (44) is to be understood as the product of eigenvalues of $(\mu^{-2} \widehat{F})$ corresponding to periodic eigenvalues only.\\

\noindent
To proceed further, we note that the functions $v(\vec x, p, \alpha) = \frac{1}{\beta} e^{i\frac{2\pi p}{\beta}\tau} u(\vec x, p, \alpha)$ are the normalized periodic eigenfunctions of $\widehat{F}$, where $u(\vec x, p, \alpha)$ are the normalized eigenfunctions of the operator obtained from $\widehat{F}$ by replacing $\partial_\tau $ with $i\frac{2\pi p}{\beta}$. Here, $\vec x$ denotes a point in the $(n-1)$ dimensional space while, $p$ is an integer and $\alpha$ is an arbitrary parameter. Thus denoting the corresponding eigenvalues by $\lambda(p, \alpha)$, one obtains

\be \widehat{F} v(\vec x, p, \alpha) = \lambda(p, \alpha)v(\vec x, p, \alpha),\ee
with the normalization and completeness conditions

\be \int v^\dag (\vec x, p, \alpha) v(\vec x, p', \alpha')d^n x = \delta_{pp'}\delta_{\alpha\alpha'}.\ee
\be \sum_{p}\sum_{\alpha}v(\vec x, p, \alpha) v^\dag(\vec x', p, \alpha) = \mathrm{I}.\ee
In expression (46) the integration is over $0\le\tau\le\beta$ and in expression (47) `${\mathrm I}$' stands for the continuous unit matrix with elements $\delta^i_j$. Now it is possible to express the heat kernel (propagator) of the operator $\widehat{F}$ as

\be K(\vec x,\vec x', s) = \sum_p\sum_\alpha v(\vec x, p, \alpha) v^\dag (\vec x', p, \alpha) e^{-\lambda(p, \alpha)s},\ee
so that it satisfies

\be \frac{\partial}{\partial s} K(\vec x, \vec x', s) + \widehat{F} K(\vec x, \vec x', s) = 0,\ee
together with the boundary condition

\be  K(\vec x, \vec x', 0) = {\mathrm I}.\ee
To express $\ln Z$ in terms of the heat kernel $K$ note that for infinitesimal variations in the background

\be \delta \left[\ln \det (\mu^{-2}\widehat{F})\right] = \sum_{p}\sum_{\alpha}\frac{\delta \lambda (p, \alpha)}{\lambda(p,\alpha)},\ee
and defining the functional trace of the heat kernel as

\be K(s) = \int \mathrm {T r} [K(\vec x, \vec x', s)]d^n x = \sum_{p}\sum_{\alpha} e^{-\lambda(p, s)s}, \ee
it is possible to obtain the following expression

\be \delta \int_0^{\infty} \frac{K(s)}{s} d s = - \sum_p\sum_\alpha \frac{\delta \lambda(p, \alpha)}{\lambda(p,\alpha)}.\ee
Now in view of expressions (51) and (53) one gets

\be \ln \left[\det(\mu^{-2} \widehat{F})\right] = - \int_{0}^{\infty}\frac{K(s)}{s}d s + \mathrm {constant}.\ee
Hence $\ln Z$ takes the form presented in expression (4).

\subsection{The expression (18) and (19) for $\zeta$ function and $w(\vec x)$}
For $n = 3$ expression (17) for $\zeta$ function takes the form

\be \zeta(\vec x, Z) = \frac{1}{\beta}\sum_{N = -\infty}^{\infty} \frac{\sqrt\gamma}{4\pi(Z -1)\Gamma(Z+1)} \int_0^{\infty}
(\mu^2 s)^{Z}\frac{\partial^2}{\partial s^2} \left[e^{-\left(m^2 + \frac{4\pi^2 N^2}{{\mathcal N}^2\beta^2}\right)s}\Lambda_{N}\right]
ds.\ee
For $n = 4$ expression (17) reads

\be \zeta(\vec x, Z) = \frac{1}{\beta}\sum_{N = -\infty}^{\infty} \frac{\mu\sqrt\gamma}{(4\pi )^{\frac{3}{2}}(Z - \frac{1}{2})(Z - \frac{3}{2})\Gamma(Z)} \int_0^{\infty}
(\mu^2 s)^{Z - \frac{1}{2}}\frac{\partial^2}{\partial s^2} \left[e^{-\left(m^2 + \frac{4\pi^2 N^2}{{\mathcal N}^2\beta^2}\right)s}\Lambda_{N}\right]
d s.\ee
For $n = 5$ dimensions expression (17) is integrated by parts yet again to yield

\be \zeta(\vec x, Z) = - \frac{1}{\beta}\sum_{N = -\infty}^{\infty} \frac{\sqrt\gamma}{(4\pi)^2(Z - 1)(Z - 2)\Gamma(Z+1)} \int_0^{\infty}
(\mu^2 s)^{Z}\frac{\partial^3}{\partial s^3} \left\{e^{-\left(m^2 + \frac{4\pi^2 N^2}{{\mathcal N}^2\beta^2}\right)s}\Lambda_{N}\right\}
d s, \ee
and so on. As in the case of $n =4$, the $\zeta$ function in even dimensional space-times contain $\Gamma(Z)$ in the denominator and so $\zeta$ functions for even dimensional space-time vanish for $Z = 0$. Hence the expression (18) is obtained in general. In view of expression (10) it is now possible to find the one-loop effective action $w(\vec x)$ for arbitrary dimension. Here we only exhibit the calculation of $w(\vec x)$ in dimensions $n = 4$, in which we are particulary interested. Since $\Gamma(0) = \infty$, so $\frac{d\zeta}{d Z}|_{Z=0}$ retains derivatives of $\frac{1}{\Gamma(Z)}$ terms and other terms vanish. Now since

\be \frac{1}{\Gamma(Z)} = Z e^{\gamma Z}\Pi_{m=1}^{\infty}\left(1+\frac{Z}{m}\right)e^{-\frac{Z}{m}}.\ee
Therefore,

\be \frac{d}{d Z}\left(\frac{1}{\Gamma(Z)}\right) = e^{\gamma Z}\Pi_{m=1}^{\infty}\left(1+\frac{Z}{m}\right)e^{-\frac{Z}{m}}\Big|_{Z=0} = 1.\ee
Since other terms contain products with $Z$, so all vanish for $Z = 0$. Hence,

\be \frac{d\zeta(\vec x, 0)}{d Z} = w(\vec x) = \frac{1}{\beta}\sum_{N = -\infty}^{\infty} \frac{\sqrt{\gamma (\vec x)}}{6(\pi)^{\frac{3}{2}}} \int_0^{\infty}
(s)^{-\frac{1}{2}}\frac{\partial^2}{\partial s^2} \left[e^{-\left(m^2 + \frac{4\pi^2 N^2}{{\mathcal N}^2\beta^2}\right)s}Tr\Lambda_{N}\right]
d s. \ee

\subsection{Explicit calculation of expression (41) for $\ln Z$ in Schwarzchild background}
\[ \ln Z = \int_0^{\beta} d\tau\int w(\vec x)d^n x\]
\be= \frac{1}{\beta}\sum_{N = -\infty}^{\infty}\int_{s=0}^{s=\infty} \frac{1}{12\pi^{\frac{3}{2}}}\int_0^{\beta}d\tau\int_0^{\pi}\sin\theta d\theta\int_0^{2\pi}d\phi\int_{2M}^{r_0}\frac{r^2}{\sqrt{1-\frac{2M}{r}}}
s^{-\frac{1}{2}}\frac{\partial^2}{\partial s^2} \left[e^{-\left(\frac{4\pi^2 N^2}{\left(1-\frac{2M}{r}\right)\beta^2}\right)s}\right]
d s dr.\ee
The Space integration has to be performed first which results in

\be \ln Z = \frac{1}{3\sqrt \pi}\sum_{K = -\infty}^{\infty}\int_{s=0}^{s=\infty} s^{-\frac{1}{2}}\frac{\partial^2}{\partial s^2} \left[\int_{2M}^{r_0}\frac{r^2}{\sqrt{1-\frac{2M}{r}}}~ e^{-\frac{K}{\left(1-\frac{2M}{r}\right)}} dr\right]d s.\ee
where, $K = \frac{4\pi^2 N^2 s}{\beta^2}$. Now to evaluate the $r$ integration, it is convenient to make the following change of variable
\be 1-\frac{2M}{r} = (Y-1)^{-1}\ee
so that the $r$ integral takes the form
\be I' = 8M^3 I = 8M^3 \int_{Y_0}^{\infty} \frac{(Y+1)^{5/2}}{Y^4} e^{-k(Y+1)} d Y, \\Y_0 = \frac{2M}{r_0 - 2M}.\ee
Clearly the integral $I$ vanishes in the limit $X \rightarrow \infty$ due to the presence of the exponent. Further, since $Y_0 \gg 1$, for $r_0 \gg 2M$, so $(Y+1)^{5/2}$ may be expanded in the power series of $Y$ and integration may be carried out term by term as follows.
\be I = e^{-K}\int_{Y_0}^{\infty}\frac{1}{Y^4}\left[1 + \frac{5}{2}Y + \frac{15}{8}Y^2 + \frac{5}{16}
Y^3 - \frac{5}{128}Y^4+...\right]e^{-KY}d Y.\ee
Now repeated integration by parts of different terms yield,
\be \int_{Y_0}^{\infty}\frac{e^{-KY}}{Y^4} d Y = \left[\frac{1}{3Y_0^3} - \frac{K}{6 Y_0^2} + \frac{K^2}{6 Y_0} + \frac{K^3}{6}\ln {Y_0}\right]e^{-K Y_0} - \frac{K^4}{6}\int_{Y_0}^{\infty}\ln{Y} e^{-KY}d Y.\ee
\be \frac{5}{2}\int_{Y_0}^{\infty}\frac{e^{-KY}}{Y^3} d Y=\frac{5}{4}\left[\frac{1}{Y_0^2} - \frac{K}{Y_0} -K^2\ln {Y_0}\right]e^{-K Y_0} +\frac{5}{4} K^3\int_{Y_0}^{\infty}\ln{Y} e^{-KY}d Y.\ee
\be \frac{15}{8}\int_{Y_0}^{\infty}\frac{e^{-KY}}{Y^2} d Y=\frac{15}{8}\left[\frac{1}{Y_0} + K\ln {Y_0}\right]e^{-K Y_0} -\frac{15}{8} K^2\int_{Y_0}^{\infty}\ln{Y} e^{-KY}d Y.\ee
\be \frac{5}{16}\int_{Y_0}^{\infty}\frac{e^{-KY}}{Y} d Y=-\frac{5}{16} \ln {Y_0}e^{-K Y_0} +\frac{5}{16} K\int_{Y_0}^{\infty}\ln{Y} e^{-KY}d Y.\ee
\be -\frac{5}{128}\int_{Y_0}^{\infty}e^{-KY} d Y=-\frac{5}{128} e^{-K Y_0}.\ee
Since higher order terms from and above fifth do not diverge ar $r_0 \rightarrow \infty$, so these are not under consideration. Therefore
\be I' = 8M^3\left[\frac{1}{3Y_0^3} - \left(\frac{5}{4} -\frac{K}{6}\right)\frac{1}{Y_0^2} + \left(\frac{K^2}{6}-\frac{5K}{4}+\frac{15}{8}\right) \frac{1}{Y_0} + \left(\frac{K^3}{6}-\frac{5K^2}{4}+\frac{15K}{8}-\frac{5}{16}\right)\ln {Y_0}\right]e^{-K Y_0 +1} + ....\ee
Therefore, considering $a_{N0}$, $\ln Z$ is expressed as
\[ \ln Z = \frac{8M^3}{3\sqrt \pi}\sum_{N=-\infty}^{\infty}\int_0^{\infty} s^{-\frac{1}{2}}\frac{\partial^2}{\partial s^2}\]
\be\left[\frac{1}{3Y_0^3} - \left(\frac{5}{4} -\frac{K}{6}\right)\frac{1}{Y_0^2} + \left(\frac{K^2}{6}-\frac{5K}{4}+\frac{15}{8}\right) \frac{1}{Y_0} + \left(\frac{K^3}{6}-\frac{5K^2}{4}+\frac{15K}{8}-\frac{5}{16}\right)\ln {Y_0}\right]e^{-K Y_0 +1} + ....\ee
To get explicit form of $\ln Z$, now one has to integrate over $s$, term by term to obtain,
\be \int_{0}^{\infty}s^{-\frac{1}{2}}\frac{\partial^2}{\partial s^2}\left[\frac{1}{3Y_0^3}+\frac{5}{4Y_0^2}+\frac{15}{8Y_0}-\frac{5}{16}\ln Y_0\right]e^{-K(Y_0+1)}ds=\sqrt{\pi(Y_0+1)^3} q^{\frac{3}{2}} \left[\frac{1}{3Y_0^3}+\frac{5}{4Y_0^2}+\frac{15}{8Y_0}-\frac{5}{16}\ln Y_0\right].\ee
\be \int_{0}^{\infty}s^{-\frac{1}{2}}\frac{\partial^2}{\partial s^2}\left[-\frac{1}{6Y_0^2}-\frac{5}{4 Y_0}+\frac{15}{8}\ln Y_0\right]K e^{-K(Y_0+1)}ds= \sqrt{\pi (Y_0+1)} q^{\frac{3}{2}} \left[\frac{1}{4Y_0^2}+\frac{15}{8Y_0}-\frac{45}{16}\ln Y_0\right].\ee
\be \int_{0}^{\infty}s^{-\frac{1}{2}}\frac{\partial^2}{\partial s^2}\left[-\frac{1}{6Y_0}-\frac{5}{4}\ln Y_0\right]K^2 e^{-K(Y_0+1)}ds= \sqrt{\frac{\pi}{(Y_0+1)}} q^{\frac{3}{2}} \left[\frac{3}{24Y_0}-\frac{15}{16}\ln Y_0\right].\ee
\be \int_{0}^{\infty}s^{-\frac{1}{2}}\frac{\partial^2}{\partial s^2}[q^3s^3] e^{-K(Y_0+1)}ds= \frac{3}{8}\sqrt{\frac{\pi}{(Y_0+1)^3}} q^{\frac{3}{2}}.\ee
\be \int_{0}^{\infty}s^{-\frac{1}{2}}\frac{\partial^2}{\partial s^2}[\frac{K^3}{6}\ln Y_0] e^{-K(Y_0+1)}ds= \frac{1}{16}\sqrt{\frac{\pi}{(Y_0+1)^3}} q^{\frac{3}{2}}\ln Y_0.\ee
In the above we have chosen $q = \frac{4\pi^2N^2}{\beta^2}$. Following a little algebra and considering $\sum_{-\infty}^{\infty}N^{-3} = \zeta_R(-3) = \frac{1}{120}$, the partition function therefore reads
\[ \ln Z = \frac{1}{34560}\left(\frac{r_0}{M}\right)^3\left(1-\frac{2M}{r_0}\right)^{\frac{3}{2}}+\frac{1}{23040}
\left(\frac{r_0}{M}\right)^2\left[5\left(1-\frac{2M}{r_0}\right)^{\frac{1}{2}}+\left(1-\frac{2M}{r_0}\right)^{\frac{3}{2}}   \right]\]
\[+\frac{1}{23040}
\left(\frac{r_0}{M}\right)\left[15\left(1-\frac{2M}{r_0}\right)^{-\frac{1}{2}}+15\left(1-\frac{2M}{r_0}\right)^{\frac{1}{2}}                      +\left(1-\frac{2M}{r_0}\right)^{\frac{3}{2}}\right]\]
\be +\frac{1}{23040}
\left[\ln\left(\frac{2M}{r_0}\right)-\ln\left(1-\frac{2M}{r_0}\right)\right]
\left[5\left(1-\frac{2M}{r_0}\right)^{-\frac{3}{2}}+45\left(1-\frac{2M}{r_0}\right)^{-\frac{1}{2}}                      +15\left(1-\frac{2M}{r_0}\right)^{\frac{1}{2}}-\left(1-\frac{2M}{r_0}\right)^{\frac{3}{2}}\right]\ee
Finally, in order to find the terms that diverge for $r_0 \rightarrow \infty$, it is required to expand $(1-\frac{2M}{r_0}^n)$ terms appearing in the above expression to obtain expression (41).

\end{document}